# AI-Driven Thermal Mapping and Management in 3D Integrated Photonic Circuits

Liton Kumar Biswas[1], Katayoon Yahyaei[1], Shajib Ghosh[1], M Shafkat M Khan[1],
Himanandhan Reddy Kottur[1], Rayhane Ghane-Motlagh[2], Mahdi Nikdast[3], Navid Asadizanjani[1]
[1]University of Florida, Gainesville, Florida 32611
[2]ficonTEC Service USA Inc., FL, USA
[3]Colorado State University, Fort Collins, Colorado 80523
Email: litonkumarbiswas@ufl.edu

## Abstract

Photonic Integrated Circuits (PICs) are advancing high-performance computing, data centers, and sensing, yet three-dimensional (3D) PICs introduce critical thermal management challenges due to high-density bonding and heterogeneous materials. Traditional methods like thermal microscopes and in-package sensors yield sparse data, limiting full thermal profile visibility. This paper presents a dual-method solution combining an AI-driven thermal modeling framework with a design-based heuristic approach. The AI method integrates sparse sensor data with design layer and density information to predict multilayer temperature variations, while the heuristic approach uses localized material properties, design layout, component geometries, and sensor coordinates to refine thermal estimations in specific regions. A 2D thermal map of a 3D PIC is generated by interpolating sensor data and adjusting for local thermal resistivity using comparative analysis between design regions. The heuristic method complements the AI model, improving estimation accuracy without extensive training data. Together, these methods offer a scalable, accurate solution for real-time thermal mapping and design-time simulation, enabling reliable thermal management in next-generation 3D photonic systems.

## I. Introduction

Photonic integrated circuits (PICs) pack many optical components – lasers, modulators, detectors, waveguides – onto a single chip, enabling ultra-fast, energy-efficient data communication [1,2]. Light carries data with minimal loss, revolutionizing long-haul fiber communications and promising to "greatly expand computing power" in data centers and AI systems if on-chip optical interconnects can be realized. Silicon photonics, leveraging mature CMOS fabrication, already provides modulators, filters, and detectors that achieve hundreds of gigabits per second per channel at low energy [2]. However, packing photonics and electronics side-by-side on a flat (2D) chip has limitations. For example, co-integrating electronics and photonics on one die "freezes" the electronics at a given technology node and limits density. At high integration scale, planar PICs require numerous waveguide crossings, which introduce optical loss and crosstalk and also exhaust chip area.

To overcome these limits, three-dimensional (3D) heterogeneous integration is emerging as the next frontier. In 3D PICs, multiple functional layers – electronic circuits, optical sources or gain layers, passive waveguide layers, etc. – are vertically stacked and coupled [3]. This stacking enables much higher device density and new functionality, for example, 3D space-division multiplexing and beam steering. Early 3D photonic-electronic systems have already demonstrated interconnect energy below 200 fJ/bit, leveraging separate optimized chips bonded together [2]. By separating the electronic driver (on an advanced CMOS chip) from the photonic layer, designers can use the latest transistors while using silicon or III–V materials optimally for optics. Such heterogeneous 3D architectures promise orders of magnitude more optical channels and modes than planar PICs, with tighter integration and co-packaged optics.

However, stacking layers with diverse materials and devices also creates a critical thermal management challenge. Every active photonic and electronic component generates heat, and in a dense 3D stack this heat must flow through thin layers of silicon, dielectrics and metals. The resulting temperature gradients can shift resonant wavelengths, reduce efficiency, and induce thermal crosstalk between channels. In what follows, we review the thermal challenges of 3D PICs and argue that AI-driven thermal mapping can offer new solutions.

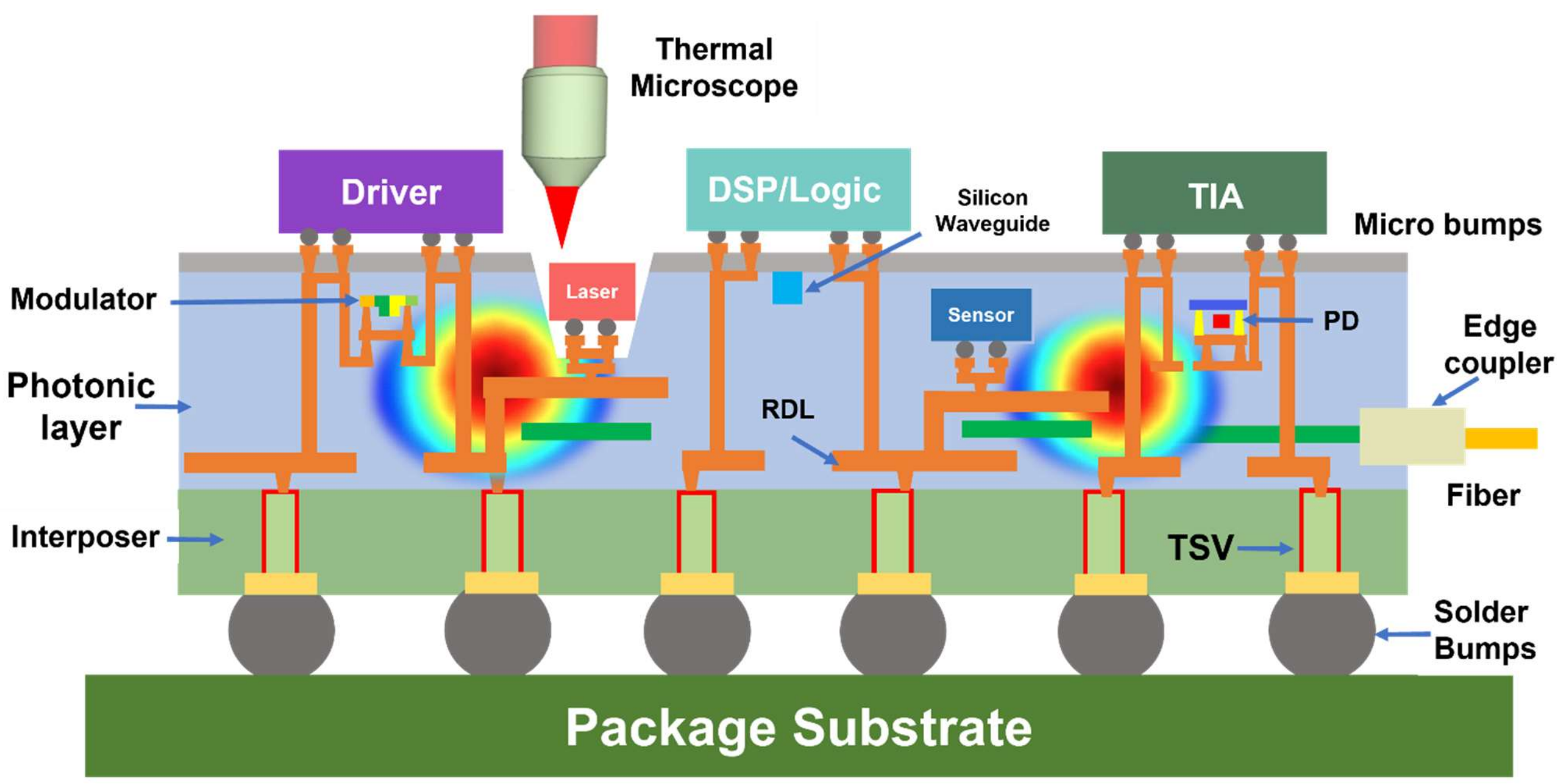

*Fig. 1: 3D Heterogeneous integration of photonics chip, showing sensors and thermal issues in structures, and thermal microscope on top*

## II. Background

*A. Planar PIC limitations and the drive to 3D:*

Conventional silicon PICs are fabricated on silicon-on-insulator (SOI) wafers, integrating waveguides, modulators, and detectors in one plane [4]. As component counts grow (for example, phased-array antennas or optical neural processors with tens of devices), planar layouts force many waveguide crossings and long interconnects. These crossings incur optical loss and limit scalability [1]. Moreover, placing electronics alongside photonics on the same chip constrains electronic performance to the photonics node, and vice versa. This has motivated heterogeneous approaches where photonics and electronics are fabricated on separate wafers (possibly using different materials) and then bonded or assembled.

In 3D heterogeneous integration (3D-HI), for example, one can bond a thin III–V die (with lasers and photodetectors) onto a silicon substrate, and then stack silicon nitride (SiN) or lithium niobate layers on top for passive routing. A representative layered 3D-PIC architecture might include (from bottom to top): a silicon CMOS electronic chip, a wafer-bonded III–V gain layer, a SiN waveguide layer, and a sapphire or silicon substrate. Electrical interconnects between layers are provided by vertical vias or bonded metal traces, while optical coupling can occur via evanescent couplers or specially fabricated vertical couplers (e.g. spiral or grating couplers) [5]. This vertical stacking unlocks a "new spatial degree of freedom" – designers can route signals not just laterally but through the depth of the chip.

As a result, 3D PICs can pack far more components in the same footprint. For example, multilayer PICs have demonstrated 3D optical phased arrays and ultradense switch fabrics. Hybrid integration also enables mixing material systems: silicon offers compact waveguides in the telecom band, while wide-bandgap layers (e.g. silicon nitride, aluminum nitride) provide low-loss routing from visible to infrared. Heterogeneous 3D PIC platforms have been developed with stacked silicon nitride and aluminum nitride waveguides on sapphire, supporting UV–NIR operation. Vertical coupling schemes (e.g. self-rolled micro-ring couplers) have been proposed to link dissimilar layers with minimal loss and fabrication complexity. These innovations demonstrate that 3D photonic integration is essential for future high-performance microsystems, but they also highlight the complexity of design and the importance of managing non-electrical constraints (like heat).

*B. Heterogeneous stacking methods:*

Three-dimensional PICs can be realized by monolithic growth (epitaxial deposition of one material on another) or by hybrid assembly (bonding prefabricated layers). For instance, silicon-on-III-V bonding can integrate a laser gain layer, while silicon nitride layers can be stacked using wafer bonding or sacrificial layer techniques. Hybrid bonding (using oxides or adhesives) is often preferred because it avoids lattice-mismatch issues. Fabrication flows typically involve planarizing layers with $SiO_2$ cladding and then patterning each layer individually. Figure 1 illustrates a simplified 3D-PIC stack on an SOI substrate: an electronic driver layer (bottom) is bonded to a silicon waveguide layer (middle) and a top photonic device layer, with vertical vias and metal interconnects linking them. (In our case study below, we consider a similar three-layer SOI platform with laser and modulator heat sources.) This schematic highlights how devices in different layers can be co-placed to achieve compact routing.

*C. Thermal considerations:*

Heterogeneous 3D integration brings materials with different thermal conductivities and expansion coefficients into close proximity. For example, silicon nitride and aluminum nitride each have very different thermal conductivities ($Si_3N_4 \approx 30$ W/m·K vs. AlN $\approx 140$ W/m·K) and thermo-optic coefficients ($4.7\times10^{-5}$/K vs. large electro-optic effect). These differences mean heat generated in one layer can spread unevenly through the stack, and local hotspots can create large temperature gradients. In typical photonic devices, even a few milliwatts of power can shift resonant wavelengths by picometers/K, so a 10–20 K rise can derail a narrowband filter. Moreover, on-chip metal interconnects (for modulators or electronics) add additional Joule heating. Finally, 3D stacking often uses high-density bonding (or eutectic), which can introduce thin layers of epoxy or metal with poor thermal conductivity [6], further complicating heat flow.

In summary, modern PICs are moving toward high-density, heterogeneous 3D architectures to meet bandwidth and integration demands. This densification leads to unprecedented thermal loads and gradients. While conventional techniques (planar circuits, IR cameras, sparse sensors) suffice for simple PICs, they face serious limitations in 3D stacks. In the next section, we discuss why thermal mapping and management in 3D-PICs are uniquely challenging and how emerging AI methods can help.

## III. Thermal Mapping Analysis Challenges

3D photonic stacks pose multiple thermal challenges that complicate performance and reliability. First, heat sources are distributed in three dimensions. Lasers, modulators, and electronic drivers (often implemented with microheaters) dissipate power in interior layers that are insulated by low-conductivity cladding and bonding. As an example, a thin SiN waveguide layer buried under oxide may host integrated heaters or amplifiers; the heat must pass through oxide and silicon before reaching the chip surface. Second, different materials conduct and store heat differently. A hotspot in a high-conductivity layer (e.g. silicon substrate) will spread heat widely, whereas in a low-κ layer (e.g. polymer or adhesive) it remains localized. Third, thermal crosstalk becomes severe: heat from one component can influence neighbors both laterally and vertically. Experimental studies have shown that even millimeter-scale separations are insufficient to isolate thermal effects in dense PICs. In one analysis of a two-layer silicon PIC with microring heaters, heating on one ring measurably shifted resonance in nearby rings several µm away, and the authors concluded that "in large thermally actuated photonic circuits, the thermal cross-talk is an issue". In 3D stacks, cross-layer crosstalk further entangles the temperature field: a heater in the top layer will raise temperature in the middle and bottom layers, perturbing devices there.

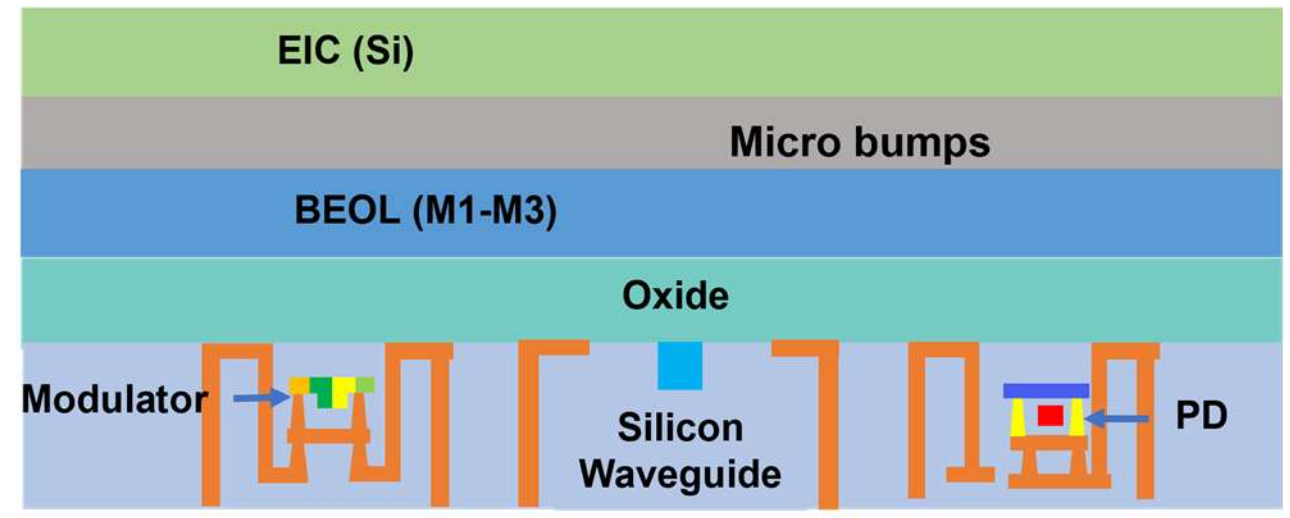

*Fig. 2: Cross-section of heterogeneous integration stack of multilayer materials of 3D photonics chip*

Accurate thermal mapping (i.e., knowing the 3D temperature distribution) is thus critical, but traditional tools fall short in 3D-PICs:

- Infrared (IR) microscopy: IR cameras can image surface temperatures by detecting thermal radiation, but the spatial resolution is limited (typically a few micrometers at best) and the method only sees the top surface [7]. In practice, buried heaters and waveguides have no direct IR signature. Moreover, emissivity variations (e.g. metals, dielectrics) make calibration difficult. As in medical thermography, "2D IR images reflect only the surface temperature distribution, hence only rough localization of the heat source is possible". For a 3D PIC, this means IR can miss or blur hotspots entirely.
- Sparse on-chip sensors: Designers sometimes incorporate discrete temperature sensors (e.g. diode or resistance thermometers) at a few points. But these provide only point readings; reconstructing a full map from a handful of probes is an underdetermined problem. Placing too many sensors is impractical (they consume area, power, and complicate the design). Moreover, any embedded sensor perturbs the local heat flow. Thus, in-package or on-chip measurements typically yield very sparse data (often just average junction temperatures), which is insufficient to capture fine gradients in a multi-layer stack.
- Conventional simulation tools: Detailed finite-element or finite-volume simulation of heat conduction can predict the temperature field, but full 3D thermal simulation is computationally expensive, especially if materials and sources are updated frequently (e.g., during circuit design optimization). Simulating the non-steady-state thermal response (transient cooling or heating) adds further complexity. In a design loop, it is often infeasible to run a full simulation at each iteration.

# IV. Case Studies for Data Collection

## A. Design Description

The hybrid 3D ring-based PIC under study features a vertically stacked architecture (in Figure 2) where a silicon photonic layer with modulator and waveguides forms the base, with additional bonded layers above. For example, the bottom Si waveguide layer (with Si core waveguides clad in $SiO_2$), while a second layer of $Si_3N_4$ or III–V material is bonded on top to implement lasers or amplifiers. Between layers, low-k dielectrics (e.g., $SiO_2$ or BCB) and oxide claddings separate the materials. Temperature sensors (e.g. embedded resistive or diode thermometers) are distributed at key locations, such as on the surface of each layer near active devices.

Silicon photonic layer: High-index Si waveguides (silicon k≈150 W/m·K) serve as the core devices. These are patterned on a silicon-on-insulator (SOI) substrate with a buried oxide ($SiO_2$, k≈1.4 W/m·K) [8].

Bonded layer: On top of the Si layer lies a bonded layer of either silicon nitride ($Si_3N_4$, k≈20–30 W/m·K) for passive routing or a III–V epitaxial film (e.g. InP, k≈68 W/m·K) for lasers/amplifiers. Active devices on this layer typically require a thick dielectric (e.g. 2–10 µm BCB) for planarization, which significantly increases thermal resistance. For instance, flip-chip InP DFB lasers on Si with a 2 µm BCB layer exhibit thermal resistances on the order of 100–200 K/W.

Dielectrics and metals: Inter-layer dielectrics ($SiO_2$) have low thermal conductivity, creating vertical thermal resistance. Metal layers (heater wires, interconnects) are highly conductive (e.g. W ≈170 W/m·K, Al ≈237, Au ≈314 W/m·K) and dominate lateral heat spreading. Prior work shows that the total heater area and metal properties are the main determinants of thermal behavior [9].

Sensors: A sparse set of thermal sensors is placed on-chip to monitor temperatures. Because of area constraints, typically only a handful of sensors can be deployed. These might be thin-film resistance thermometers or diode-based sensors patterned near the heated up areas, providing localized temperature readings.

Key thermal properties of the materials are summarized above: silicon's high k allows efficient in-plane conduction, whereas $SiO_2$ cladding and thick bonding oxides act as insulators. The heterogeneous stack, therefore, creates a complex thermal landscape. In particular, silicon nitride and III–V layers have much lower k than Si, and each material interface adds thermal boundary resistance. The thick oxide cladding common in Si-photonics further hinders vertical heat flow. These diverse materials (Si, $SiO_2$, $Si_3N_4$, III–V, metals) with mismatched conductivities and expansion coefficients make accurate thermal modeling critical.

## B. Experimental requirements

A feasible proof-of-concept experiment would proceed as follows:

*Prototype 3D PIC test chip:* A small 3D photonic chip containing several waveguides on the Si layer needs to be fabricated and, if possible, a bonded layer (e.g. InP or SiN) on top. Integrate thin-film heater traces (W or Al) on each for thermal tuning, and co-fabricate a limited number of temperature sensors (e.g. metal resistance thermometers) at strategic locations.

*Sensor readout and calibration:* Each on-chip sensor to an external readout (microcontroller or DAQ) to be connected, and can be calibrated them using a known temperature reference. Optionally, a calibrated IR microscope to measure surface temperatures can be employed (although the thick oxide may limit IR penetration).

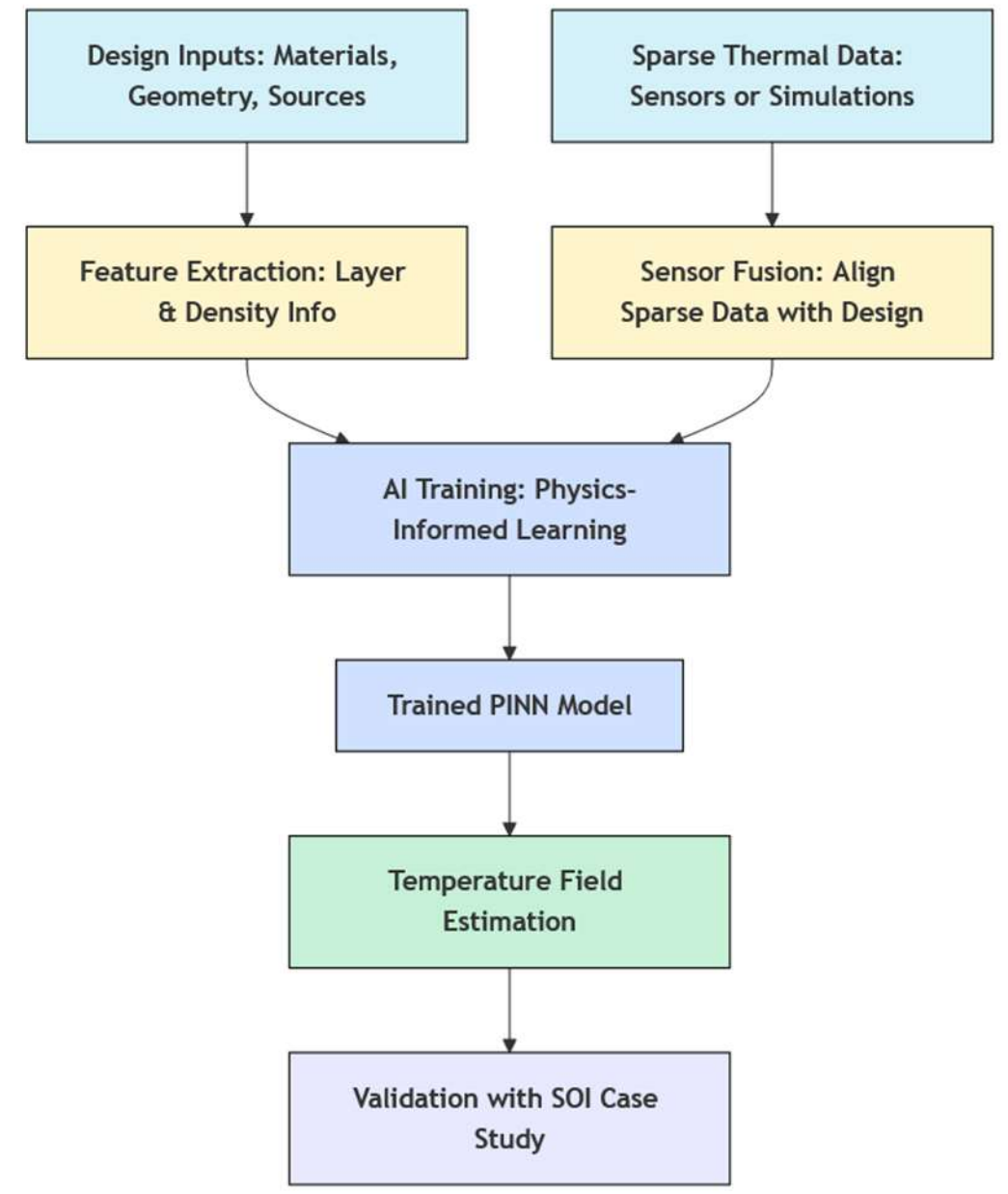

*Fig. 3: AI-based thermal estimation framework for 3D photonic integrated circuits (PICs).*

*Controlled heating experiments:* Sequentially activate ring heaters with known currents. For example, apply step currents (e.g. 0–20 mA) to one heater at a time, corresponding to ~0–30 mW dissipation. Record the resulting temperature at each sensor over time to capture both steady-state and transient responses. Repeat for different heaters, combinations of heaters, and various power levels.

*Data sparsity:* During measurements, use only the on-chip sensors (3–5 points) to simulate sparse data. For each heating condition, collect the sparse sensor outputs together with a reference temperature map (from the IR image or the known test pattern).

## V. Thermal Profiling Analysis

### *A. AI-based Framework*

The proposed AI framework estimates the complete three-dimensional (3D) thermal distribution within a multilayer photonic integrated circuit (3D PIC) by utilizing sparse thermal measurements alongside comprehensive design metadata. It initiates with two main inputs: (1) design details—comprising material stack, geometric configuration, interconnect density, and identified heat sources—and (2) sparse thermal data gathered from simulations (for instance, silicon waveguides on SOI) or embedded sensors. The design inputs undergo processing through a feature extraction module that encodes thermally relevant characteristics such as material boundaries, conductivity contrasts, and spatial distributions of heat sources. Concurrently, the sparse temperature measurements are synchronized with the design layout through a sensor fusion module, ensuring spatial coherence and facilitating partial supervision.

These processed data streams are then directed into the AI training module, where a Physics-Informed Neural Network (PINN) is trained. In contrast to black-box models, the PINN incorporates the physical heat conduction equation $\nabla\cdot(k\nabla T) = Q$ directly into its loss function, thereby enforcing thermodynamic consistency during the learning process. The training process utilizes both sparse empirical data and established physics principles, with the option to integrate hybrid-fidelity data sources—such as coarse analytical solutions and detailed finite-element (FEM) simulations. Once the model is trained, it produces a high-resolution 3D temperature field, even in areas lacking measurements. The inferred results undergo validation against a simplified experimental setup, like a silicon waveguide on SOI, to showcase accuracy and practicality. This data-efficient, design-sensitive, and physically grounded framework facilitates scalable thermal analysis of next-generation 3D PICs without the need for exhaustive simulation or extensive sensor coverage.

### *B. Heuristic design-based Analysis*

Design-based thermal analysis is proposed due to the significance of design information in understanding and addressing heat dissipation, thermal resistivity, and local hotspot formation in complex systems. The heuristic design-based approach generates a 2D thermal profile of a 3D PIC by considering localized design details, such as the specific materials used in a given cross-sectional area and the precise thermal data captured by integrated sensors. This method provides a granular estimation of temperature variations with focus on localized design information, examining small areas within the cross-section (e.g., between coordinates ($x_1$, $y_1$) and ($x_2$, $y_2$) shown in Figure 4) and considering the materials used in those specific regions. This localized approach allows for more accurate thermal data estimation in these areas by referencing nearby locations.

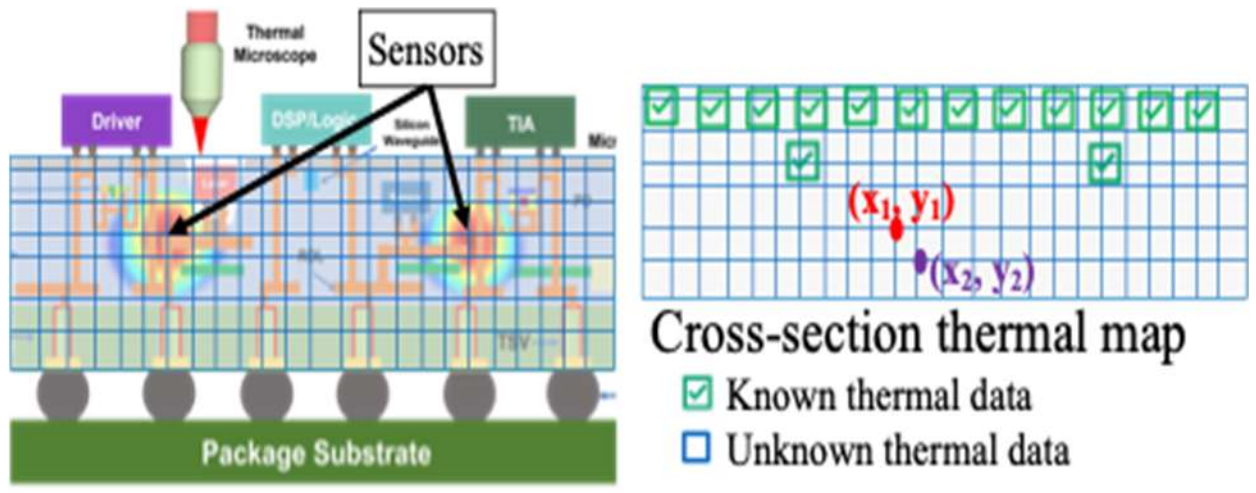

*Figure 4: Illustration of thermal map grid processed from 3D PIC design*

The proposed approach, as shown in Figure 5, utilizes sensor thermal data, the sensor's coordinates within the 3D space of the package, and detailed, coordinated layout and design information. When available, thermal microscope data enhances accuracy by providing additional known thermal points on the surface. As illustrated in Figure 4, the input sensor thermal and spatial data are used to create a grid representing the desired 2D thermal profile, specifying known thermal data points. A heuristic estimation is then performed for grid cells with unknown thermal data. The simplest form of this estimation is linear interpolation, which assumes a linear distribution of temperature data across the grid cells. To refine these estimations, localized design details within each grid cell are factored in. By conducting a comparative analysis of thermal conductivities and resistances with neighboring cells, the method adjusts the initial linear interpolation results. This adjustment process ensures that the thermal estimations account for the material-specific properties and spatial variations, thereby improving the overall accuracy of the thermal profile.

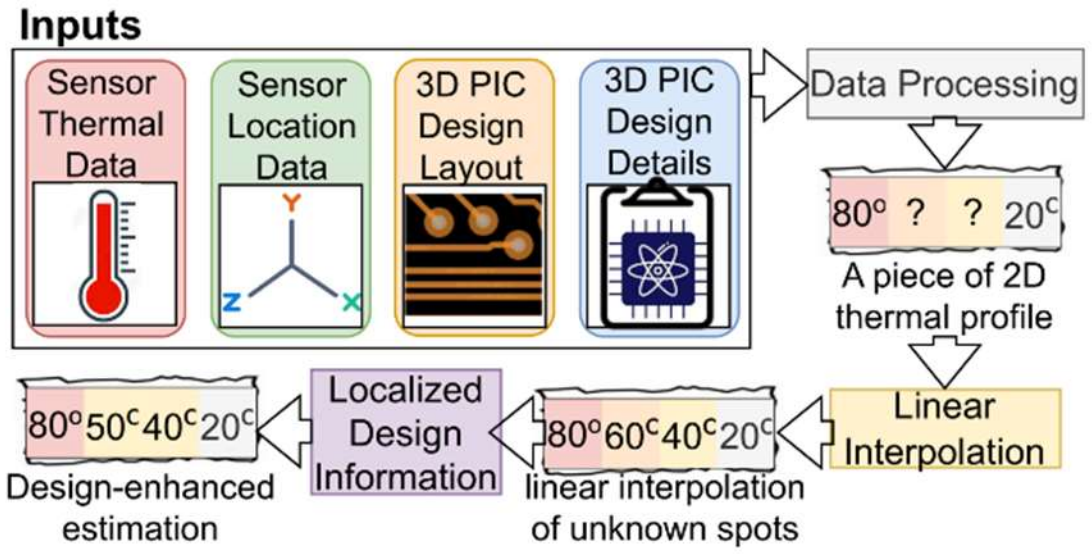

*Figure 5: Workflow of heuristic design-based approach*

The AI-based and heuristic design-based thermal analysis methods, while both relying on design specifications and sensor data, differ in their scope and approach. The AI-based method processes holistic design information through feature extraction and a PINN to model broad thermal behavior, but it requires an initial training dataset, which can be challenging to obtain. Conversely, the heuristic design-based approach focuses on localized circuit areas, using detailed design data to estimate thermal properties with greater precision and operating independently with minimal or no prior data. Despite these differences, the heuristic method can complement the AI-based framework, validating and refining its outputs to ensure estimates remain within acceptable ranges when direct validation is unavailable. This synergy between the two methods enhances the overall accuracy and reliability of thermal analyses in complex systems.

## VI. Conclusion

This work addresses the critical thermal challenges in 3D PICs by introducing a hybrid solution that combines AI-driven modeling with a heuristic design-based approach. The AI framework utilizes sparse sensor and design data to predict complex thermal behavior across multilayer architectures, while the heuristic method leverages localized material properties and spatial layout to enhance estimation accuracy in specific regions. By integrating these complementary approaches, we establish a robust strategy for both design-base simulation and real-time thermal monitoring. The design-based method offers independence from large training datasets and enables precise thermal interpolation where direct measurements are sparse. Together, these methods provide a scalable and adaptable solution for comprehensive thermal analysis, supporting the development of thermally reliable and efficient 3D-PICs. Future work will explore tighter coupling between AI and heuristic models to further improve prediction robustness and adaptability across diverse packaging configurations.